\documentclass[10pt,letterpaper]{article}
\usepackage[top=0.85in,left=1.5in,footskip=0.75in]{geometry}

\usepackage{amsmath,amssymb}

\usepackage{changepage}

\usepackage{textcomp,marvosym}

\usepackage{cite}

\usepackage{nameref,hyperref}

\usepackage[right]{lineno}

\usepackage[nopatch=eqnum]{microtype}
\DisableLigatures[f]{encoding = *, family = * }

\usepackage[table]{xcolor}

\usepackage{array}

\newcolumntype{+}{!{\vrule width 2pt}}

\newlength\savedwidth

\raggedright
\usepackage[aboveskip=1pt,labelfont=bf,labelsep=period,justification=raggedright,singlelinecheck=off]{caption}

\makeatletter
\renewcommand{\@biblabel}[1]{\quad#1.}
\makeatother

\usepackage{lastpage,fancyhdr,graphicx}
\usepackage{epstopdf}
\fancyheadoffset[L]{1.25in}
\fancyfootoffset[L]{1.25in}
\begin{document}
\vspace*{0.2in}

\begin{flushleft}
{\Large
\textbf\newline{Extraction of Human DNA from Soil: Protocol Adaptations} 
}
\newline
\\
Wera M Schmerer
\\
\bigskip

\textbf{}Department of Biology, Chemistry and Forensic Science, School of Lifescience, University of Wolverhampton, Wolverhampton, UK
\\
\bigskip

w.schmerer@wlv.ac.uk

\end{flushleft}

\section*{Key words}
DNA extraction, soil samples, humic acids, PCR inhibition, protocol adaptation, human identification

\section*{Abstract}
PCR-based analysis of DNA is utilized in a wide variety of fields, including Forensic Science. Aside from the more common ample sources, material analyzed here can refer to specimen excavated from a soil environment, or a sampling of the soil itself to recover DNA leached into the soil from decomposing human remains or from body fluids intermingled with the soil in an outdoor crime scene.

The common problematic of these types of sample is the presence of humic acids, which are a component of any soil environment, and when the co-extracted with the DNA, lead to inhibition of enzyme-based procedures including PCR.

While a variety of methods exist for the extraction of DNA from excavated skeletal remains, protocols for extraction of DNA from the soil directly are usually targeting soil microorganism.
To address the need for methodology suitable for extraction of human DNA from soil, a selection of three published protocols were adapted for this purpose, to be tested and evaluated using standardized samples. The resulting protocols are presented here.

\section*{1 Introduction}
PCR-based analysis of Deoxyribonucleic Acid (DNA) is common methodology in a wide variety of fields, including Forensic Science. Here, the approach is predominantly utilized to identify individuals or trace biological samples back to their source individual.
In tis effort, a wide variety of sample material is analyzed, including soil samples with traces of human DNA, respectively DNA-containing body fluids.

During the decomposition process, DNA is leaching into the ground from the decomposing body (Emmonds 2015, Thomas et al 2019). Crime scenes in an outdoor setting may include body fluids such as blood intermingled with soil as evidence (Shahzad et al 2009, Batu-Boateng et al 2018). In either case, DNA can survive in the soil environment, even as free extracellular DNA bound to soil minerals (Willerslev et al 2003).
\paragraph{}
Samples of soil, or any material recovered from soil, such as excavated skeletonized human remains, will inevitably be affected by the presence of humic substances (humic acids), which are potent enzyme inhibitors (Pääbo et al 1989, Kraeder 1996, Wilson 1997, Aleddini 2012) and may co-extract with the DNA during extraction procedures (e.g. Hänni et al 1995). Where such occurs, enzyme dependent analyses of DNA will be inversely affected due to the inhibitory effect of humic acid (e.g. Pääbo et al 1989, Primorac et al 1996, Kraeder 1996).
\paragraph{}
Skeletal remains in a (soil) burial, will be subject to physico-chemical interaction and exchange with the soil environment over the burial interval (Herrmann \& Nesley 1982, Pate et al 1989), one effect of which is the influx of humic substances from the soil into the bone tissue. That such occurs even after a short period in a soil environment, can be shown by histological analysis of the bone tissue (Schmerer 1999, 2000). The speed of this influx, and thus the extend at any given point in time, may differ between soil environments, dependent on their composition and characteristics (Schmerer 1999, 2000), but it will occur in each case, since humic substances are an essential component of the organic matter portion of any soil (e.g. Schnitzer 1978, Tan 2014, Ukalska-Jaruga et al 2021).

In consequence, the presence of humic acids and co-extraction of these during DNA extraction procedures resulting in inhibition of analyses is a generalized problematic where samples of soil or matter recovered from soil is analyzed by PCR-based methodology. This emphasizes the need for optimized protocols to minimize the co-extraction of humic acids with the DNA. 
\paragraph{}
To this point, there is a lack of published protocols optimized for the extraction of human DNA having leached into the soil or is contained in body fluids mixed with soil (Howarth et al 2022). 
The majority of existing protocols are designed for extraction of DNA from the bacterial community in the soil (e.g. Tsai \& Olson 1991) or optimized for extraction of DNA contained in skeletal remains recovered from soil burials (e.g. Hänni et al 1990, 1995, Schmerer et al 1999, Schmerer 2003, 2021a), and thus optimized for a different species of sample material, with its sample specific requirements.

Experimental experience shows, that a protocol can successfully be adapted to be utilized for extraction of DNA for a variety of source materials, provided sample-specific requirements are appreciated (Schmerer 2021a, 2021b). Such usually entails adaptation of the sample preparation procedures prior to the actual extraction process to the specific requirements of a given sample material (see e.g. Walsh et al 1991, Schmerer 2022).
\paragraph{}
The aim of the study here is to develop and evaluate modified versions of published protocols, which were adapted for the extraction of human DNA from soil samples.
Presented here are the outcomes of the protocol adaptation.

\section*{2 Adaptation of published protocols}
Included in the adaptation process were three protocols, each of which represents one of the major approaches to DNA extraction; organic solvent- (phenol-chloroform), resin- (Chelex 100) and silica matrix-based procedures. 

Based on established procedures for similar applications (Willerslev et al 2003), protocols were adapted to process equal amounts of 0.25g of soil. With reference to the findings by Howarth et al. (2022), steps to retrieve the DNA from the soil matrix were carried out uing phosphate buffered saline (PBS) to stabilize the DNA, instead of the commonly used EDTA solution (see protocols II \& III).
\subsection*{2.1 Organic solvent (phenol/chloroform) based extraction – Protocol I}
As representative for the organic extraction approach Tsai \& Olson (1991) was selected. This widely cited protocol was designed for direct extraction of microorganism DNA from soil. Consequently, a number of steps involved in this procedure are specific to retrieving microorganism despite their interaction with the soil particles and lysis of a rather tough cell wall.

Sample preparation in original procedure (Tsai \& Olson 1991) consist of the soil sample being subjected to two washes in phosphate buffer, followed by lysis of the pelleted material in the presence of NaCl, EDTA and lysozyme and three freeze-thaw cycles following addition of a NaCl, Tris-HCl, SDS mixture to release the DNA from the cell.
\paragraph{}
Where the extraction of DNA leached into the soil or body fluids intermingled with soil is intended, the initial wash steps here would result in washing out, and discarding, at least a portion of the target DNA and was thus abandoned, as was the use of freeze-thaw cycles, which are redundant in case of comparatively easily lyzable animal cell walls.
These step were replaced by direct lysis of the soil sample n the presence of a lysis buffer consisting of NaCl, Tris-HCl, EDTA and SDS in the presence of proteinase K. 
\paragraph{}
The extraction part of the Tsai \& Olson (1991) protocol continues with successive extractions using phenol, followed by a chloroform/isoamyl mixture (24:1) and then pure chloroform.

In the adapted protocol, this was replaced by two successive extractions using a phenol/chloroform/ isoamyl mixture (24:24:1), followed by chloroform (100\%), replacing the originally used reagents with updated formulations.
\paragraph{}
To purify the resulting DNA, Tsai \& Olson (1991) use a precipitation in isopropanol, vacuum drying, followed by re-suspension in TE buffer, RNase treatment and additional column-based purification.
For the purpose of analyzing human DNA, RNase treatments are not required, which then likewise removes the need for a column-based removal of associated reactants and products.
Instead, the adapted protocol utilizes a precipitation in cold isopropanol in the presence of silica solution, followed by air-drying of the pellet and re-suspension in PCR-grade water.

\subsection*{Protocol I (phenol/chloroform approach):} 

Adaptation of Tsai \& Olson (1991) to extraction of Human DNA from soil
\paragraph{}
Sample Preparation:

\begin{enumerate}
	\item{0.25g soil is sampled and placed in a 2ml reaction tube (SafeSeal, Sarstedt),}
	\item{mixed with 500µl lysis buffer (0.15M NaCl, 10\% sodium dodecyl sulfate (SDS), 0.1M Na\textsubscript{2}EDTA  [pH 8.0])}
	\item{20µl proteinase K (10mg/ml) are added, followed by thorough mixing}
	\item{The lid of the 2ml reaction tube should be sealed using a thin strip of Parafilm (Bemis Corp.)}
	\item{For cell lysis, samples are incubated at 56°C for 2h under constant agitation
(e.g. shaking waterbath)}
\end{enumerate}

Phenol/chloroform extraction:

\begin{enumerate}	
	\item{An equal volume (520µl) of phenol mixture (phenol:chloroform:isoamylalcohol in the ratio 24:24:1) is added}
	\item{The mixture should be briefly vortex to obtain an emulsion, followed by}
	\item{centrifugation at 6000rpm (ca. 3000 x g, Micro Centaur, MSE), for 10min to reach phase separation}  
	\item{The aquatic phase is carefully transferred to a new 2ml tube} 
	\item{Followed by a repetition of the phenol step by adding an equal volume of phenol mixture (steps 1.-4.)}
	\item{An equal volume (520µl) of chloroform (100\%) is added,}
	\item{proceeding as previously with the phenol mixture by repeating steps 2.-4.}
\end{enumerate}

\begin{figure}[!h]
\includegraphics[width=5in]{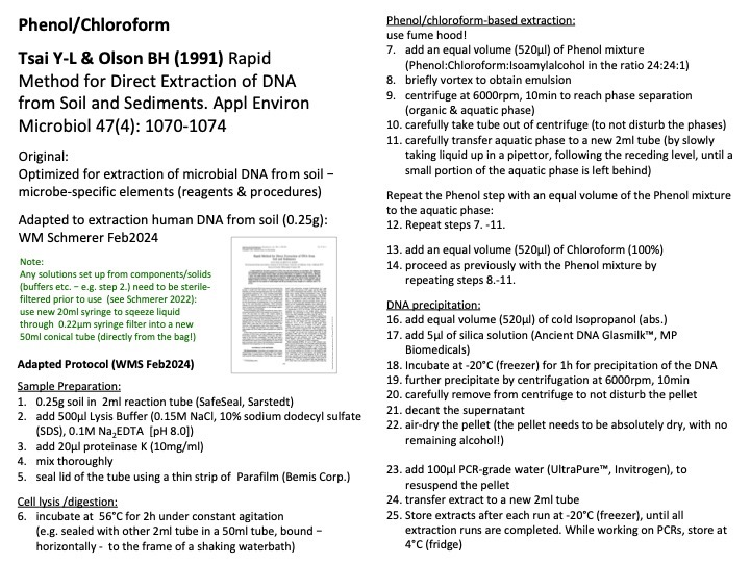}
\centering
\caption{{\bf  Protocol I - Phenol/chloroform-based protocol (Tsai \& Olson 1991), adapted for extraction of human DNA from soil (Schmerer 2024)}}
\label{fig1}
\end{figure}

DNA purification:

\begin{enumerate}	
	\item{For precipitation of the DNA, an equal volume (520µl) of cold isopropanol (abs.) and 5µl of silica solution (Ancient DNA Glasmilk™, MP Biomedicals) are added}
	\item{The mixture is incubated at -20°C (freezer) for 1h, followed by centrifugation at 6000rpm (ca. 3000 x g), for 10min}
	\item{The supernatant is decanted}
	\item{The pellet is air-dried,}
	\item{and re-suspended in 100µl PCR-grade water (UltraPure™, Invitrogen)}
	\item{Extract is transferred to a new 2ml tube}
\end{enumerate}
\paragraph{}
A more detailed version of the protocol with additional procedural instruction can be seen in Fig~\ref{fig1} 
\paragraph{}

\subsection*{ 2.2 Resin (Chelex 100) based extraction – Protocol II} 

The resin-based approach to DNA extraction is represented by the Chelex-100 based protocol by Schmerer (2021a). This protocol is already optimized for limiting the co-extraction of humic acids, but was originally developed with focus on the extraction from historical skeletal remains excavated from burials in a soil environment. Subsequent tests on a variety of sample materials showed that with material-specific modification of the sample preparation steps, this protocol can be successfully applied to a wide variety of sample materials aside from the one it was designed for, including material containing high concentrations of humic acids (Schmerer 2021a, b).
\paragraph{}
Sample preparation in original procedure (Schmerer 2021a) is specific to extraction from bone samples and consists of decontamination, pulverization, decalcification for several days in 0.5M EDTA solution, and lysis of the supernatant in the presence of proteinase K for 60-90min depending on the preservation of the sample.
\paragraph{}
To adapt this protocol to the extraction of human DNA from soil, bone matrix-specific reagents and procedures were replaced with a soil matrix specific sample preparation, the focus of which is to physically retrieve the DNA, respectively DNA-containing cells from the soil matrix, while stabilizing the DNA to prevent degradation during processing. 

For this purpose, an extended incubation in phosphate buffered saline (1x PBS) is utilized. Following centrifugation, the supernatant is subjected to the remainder of the procedure.
\paragraph{}
Since the retrieval step here utilizes a significantly smaller volume compared to the decalcification step in the original protocol, all volumes for the subsequent steps were adjusted accordingly, which allows for use of 2ml reaction tubes throughout. Aside from the reduction in volume, the adapted protocol follows the original procedure for extraction and purification of DNA.

\subsection*{Protocol II (Chelex 100 approach):} 

Adaptation of Schmerer (2021a) to extraction of Human DNA from soil
 paragraph{}
 
Sample Preparation:
 
 \begin{enumerate}	
	\item{0.25g soil is sampled and placed in a 2ml reaction tube (SafeSeal, Sarstedt),}
	\item{mixed with 500µl PBS (1X PBS: 137mM sodium chloride, 2.7mM potassium chloride, and 10mM phosphate buffer)}
	\item{The lid of the 2ml reaction tube should be sealed using a thin strip of Parafilm (Bemis Corp.)}
	\item{Sample is incubated at 40°C for 4h under constant agitation (e.g. shaking waterbath)}
	\item{Soil is pelleted by centrifugation at 6000rpm (3000 x g, Micro Centaur, MSE) for 5min} 
	\item{Supernatant is transferred to a new 2ml reaction tube}
\end{enumerate}
 
Chelex-based extraction:
 
\begin{enumerate}	
	\item{The supernatant from the previous step is mixed with 500µl Chelex-100 solution (5\% (w/v) in PCR-grade water (UltraPure™, Invitrogen),}
	\item{20µl proteinase K (10mg/ml) are added, followed by thorough mixing}
	\item{The lid of the 2ml reaction tube should be sealed using a thin strip of Parafilm (Bemis Corp.)}
	\item{Samples are incubated at 56°C for 1h under constant agitation (e.g. shaking waterbath),}
	\item{then incubated at 95°C for 8min (e.g. boiling water bath)}
	\item{Samples are left to cool to RT}
	\item{Chelex beads are pelleted by centrifugation at 4000rpm (ca. 1300 x g), 6min}
	\item{The aqueous supernatant is carefully transferred to a new 2ml tube} 
  \end{enumerate}
  
DNA purification: 
  
\begin{enumerate}
	\item{For precipitation of the DNA, an equal volume (1ml) of Isopropanol (abs., RT), 20µl
sodium acetate buffer (2M, pH 4.5) and 5µl of silica solution (Ancient DNA Glasmilk™, MP Biomedicals) are added,}
	\item{followed by incubation at RT for 30min for precipitation of the DNA,}
	\item{and centrifugation at 4000rpm (ca. 1300 x g), for 5min} 
	\item{The supernatant is decanted}
	\item{The pellet is washed in 500µl ethanol (abs.)}
	\item{air-dried, and resuspend in 100µl PCR-grade water (UltraPure™, Invitrogen)}
	\item{The resulting extract is transferred to a new 2ml tube} 
 \end{enumerate}

\begin{figure}[!h]
\includegraphics[width=5in]{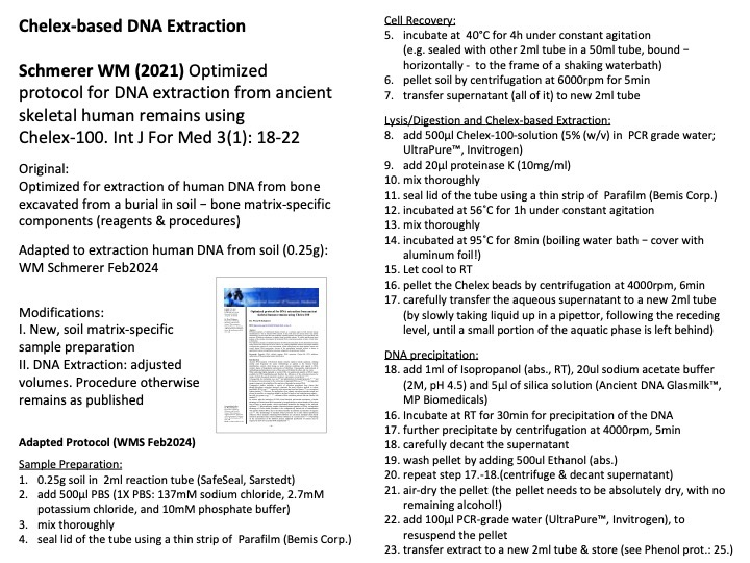}
\centering
\caption{{\bf  Protocol II – Chelex-based protocol (Schmerer2021a), adapted for extraction of human DNA from soil (Schmerer 2024)}}
\label{fig2}
\end{figure}

\paragraph{}
A more detailed version of the protocol with additional procedural instruction can be seen in Fig~\ref{fig2} 
\paragraph{}

\subsection*{2.3 Silica column based extraction – Protocol III} 

The method selected for the silica-based extraction (Schmerer 2022) is an approach for a streamlined parallel extraction of DNA from a variety of source materials, where individually timed, sample specific preparations are followed by using one common DNA extraction procedure. This approach in itself demonstrates that by utilizing a material-specific sample preparation step, a protocol may be adapted to be suitable for a variety of different sample sources.
\paragraph{}
To adapt this protocol to the extraction of DNA from human soil, the material-specific sample preparation utilized in the adaptation of the Chelex approach (Schmerer 2021a, see 2.2) was combined with the “common extraction protocol” aspect of the procedure as published, which consists of a modified version of a silica-kit based extraction protocol (Schmerer 2022). The protocol utilized is a modification of the manufacturer instruction for the E.Z.N.A.® Blood DNA Mini Kit using the "Buccal Swabs Protocol" (Omega Bio-Tek 2019); without addition of PBS to the sample, an extended proteinase K digest (30min instead of 10min), elution in the minimum volume of pre-heated PCR-grade water instead of using Elution Buffer, and with reapplication of the eluate to the silica column for a second elution, instead of using a further aliquot of water.

A pre-digestion step, as featured in the protocol overview (see Fig 3), can be included but is not essential.

\subsection*{Protocol III (Silica kit based approach):} 
Adaptation of Schmerer (2022) to extraction of Human DNA from soil
paragraph{}

Sample Preparation:

\begin{enumerate}
	\item{0.25g soil is sampled and placed in a 2ml reaction tube (SafeSeal, Sarstedt),}
	\item{mixed with 500µl PBS (1X PBS: 137mM sodium chloride, 2.7mM potassium chloride, and 10mM phosphate buffer)}
	\item{The lid of the 2ml reaction tube should be sealed using a thin strip of Parafilm (Bemis Corp.)}
	\item{Sample is incubated at 40°C for 4h under constant agitation (e.g. shaking waterbath)}
	\item{Soil is pelleted by centrifugation at 6000rpm (3000 x g, Micro Centaur, MSE) for 5min} 
	\item{Supernatant is transferred to a new 2ml reaction tube}
 \end{enumerate}
 
Pre-Digestion:
 
\begin{enumerate}
	\item{After adding 20µl proteinase K (10mg/ml), and thorough mixing}
	\item{the sample is incubated at 56°C for 1h under constant agitation (e.g. shaking waterbath or thermomixer)}
\end{enumerate}

Silica kit-based extraction:

\begin{enumerate}
	\item{After addition of 25µl Proteinase K Solution (kit component) and 500µl BL Buffer (kit component), the sample is thoroughly mixed}
	\item{Incubation at 65°C for 30min}
	\item{The proteinase K digest is terminated by addition of 100µl of Ethanol (abs.) and thorough mixing}
	\item{The sample can be briefly centrifuged to avoid liquid remaining in the lid of the tube}
	\item{A silica column (kit component: HiBind® DNA Mini Column) is placed in a 2ml collection tube (kit component)}
	\item{750µl of the sample is transferred to the column}
	\item{After centrifugation at 10,000 x g for 1min, the filtrate is discarded}
	\item{Steps 6 and 7 are repeated until the remainder of the sample has been transferred completely, after which the column is placed into a new collection tube}
	\item{Following addition of 500µl HBC Buffer (kit component), the column is centrifuged at 10,000 x g for 1min and the filtrate is discarded}
	\item{After adding 700µl DNA Wash Buffer (kit component), the column is centrifuged at 10,000 x g for 1min and the filtrate is discarded}
	\item{The wash step is repeated with a second aliquot of 700µll DNA Wash Buffer}
	\item{To remove any residual ethanol, the column matrix is dried by further centrifugation at maximum speed (at least 10,000 x g) for 2min, after which the column is transferred to a 2ml (SafeSeal, Sarsted)}
	\item{To elute the DNA from the silica matrix, 100µl PCR-grade water (UltraPure™, Invitrogen), pre-heated to 65°C, are added to the column, and incubated for 5min at RT, followed by centrifugation at 10,000 x g for 1min}
	\item{The elution step is repeated by re-applying the eluate to the column, followed by incubation and centrifugation as described in step 13, to elute any DNA remaining bound to the silica matrix after the first elution}
\end{enumerate}

\begin{figure}[!h]
\includegraphics[width=5in]{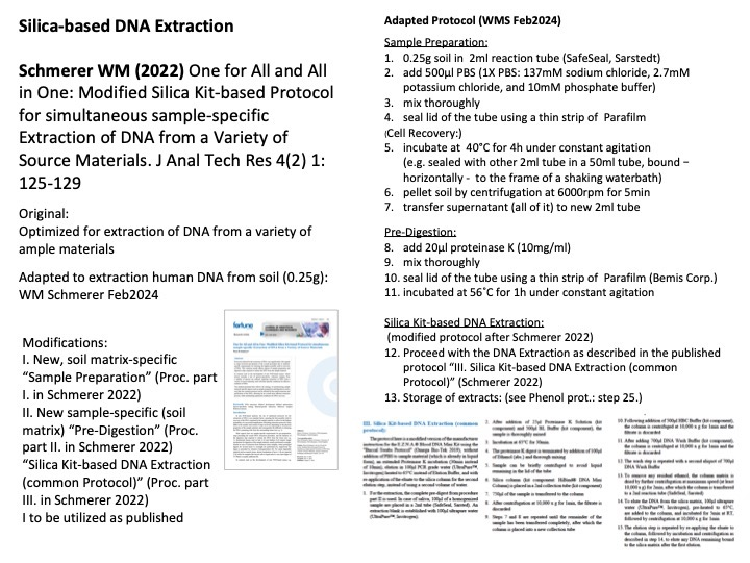}
\centering
\caption{{\bf  Protocol III – Silica column-based protocol (Schmerer 2022), adapted for extraction of human DNA from soil (Schmerer 2024)}}
\label{fig3}
\end{figure}

\section*{3. Evaluation of adapted protocols}

To test and evaluate the adapted protocols, the study utilizes standardized samples consisting of 0.25g homogenized soil mixed with identical amounts of a well-mixed saliva sample. Saliva is used as DNA source to avoid adding further inhibitory substances, as present in other DNA-containing body fluid such as blood (Wilson 1997).

Soil blanks are included to determine amounts of background DNA from the soil microorganisms. With this type of control in place, DNA yield can be determined by UV spectrophotometry, which likewise allows for quantification of the concentration of humic acids to monitor the level of co-extraction of this inhibitor. Since humic acids do not have a distinctive absorbtion band, but show increased UV absorbtion at the lover wavelenths (Kumada 1955), measurements are taken at 320nm and 360nm, to avoid the absorbtion ranges of most proteins.

Quality of the extracted DNA is evaluated by fragmentation assay (agarose gel electrophoresis of extracts) and PCR-based amplification of common STR loci (see Schmerer et al. 1999). The latter approach likewise allows for relative quantification of amplifyable DNA.

\section*{4. 4. References}
\paragraph{}
Alaeddini R (2012) Forensic implications of PCR inhibition - A review. Forensic Sci Int Genet 6: 297-305
\paragraph{}
Badu-Boateng A, Twumasi P, Salifu SP \& Afrifah KA (2018) A comparative study of different laboratory storage conditions for enhanced DNA analysis of crime scene soil-blood mixed sample. Forensic Sci Int 292: 97-109
\paragraph{}
Emmons AL (2015) The Preservation and Persistence of Human DNA in Soil during Cadaver Decomposition. MA Thesis, University of Tennessee. Available at: https://trace.tennessee.edu/utk\_gradthes/3332/
\paragraph{}
Hänni C, Brousseau T, Laudet V \& Stehelin D (1995) Isopropanol precipitation removes PCR inhibitors from ancient bone extracts. Nucleic Acids Res 23(5): 881–882
\paragraph{}
Hänni C, Laudet V, Sakka M, Begue A \& Stehelin D (1990) Amplification de fragments d'ADN mitochondrial a partir de dents et d'os humains anciens. C R Acad Sci Paris 310: 365–370
\paragraph{}
Herrmann B \& Newesely H (1982) Dekompositionsvorgänge des Knochens unter langer Liegezeit. I. Die mineralische Phase. Anthrop Anz 40: 19-32
\paragraph{}
Howarth A, Drummond B, Wasef S \& Matheson CD (2022) An assessment of DNA extraction methods from blood-stained soil in forensic science. Forensic Sci Int 341: 111502
\paragraph{}
Kreader CA (1996) Relief of Amplification Inhibition in PCR with Bovine Serum Albumin or T4 Gene 32 Protein. Appl. Environ. Microbiol 62(3): 1102–1106
\paragraph{}
Kumada K (1955) Absorption spectra of humic acids. Soil Sci Plant Nutr 1(1): 29-30
\paragraph{}
Omega Bio-Tek (2019) E.Z.N.A.® Blood DNA Mini Kit. Product Manual. January 2019. Rev.No. v8.0. Available at: www.omegabiotek.com
\paragraph{}
Pääbo S, Higuchi RG \& Wilson AC (1989) Ancient DNA and the polymerase chain reaction. The emerging field of molecular archaeology. J Biol Chem 264(17): 9709-9712
\paragraph{}
Pate FD, Hutton JT \& Norrish K (1989) Ionic exchange between soil solution and bone: toward a predictive model. Appl Geochem 4(3): 303-316
\paragraph{}
Primorac D, Andelinovic S, Definis-Gojanovic SM, Drmic I, Rezic B, Baden MM, Kennedy MA, Schanfield MS, Skakel SB \& Lee HC (1996) Identification of war victims from mass graves in Croatia, Bosnia, and Herzegovina by use of standard forensic methods and DNA typing. J Forensic Sci 41(5): 891–894
\paragraph{}
Schmerer WM (1999) Optimierung der STR-Genotypenanalyse an Extrakten alter DNA aus bodengelagertem menschlichen Skelettmaterial. Doctoral Dissertation, University of Göttingen, Germany, p 87-187
\paragraph{}
Schmerer WM (2000) Chap. 7 Degradierung von DNA im Knochengewebe in Relation zu Liegemilieu und Liegezeit. In: Optimierung der STR-Genotypenanalyse an Extrakten alter DNA aus bodengelagertem menschlichen Skelettmaterial. Cuvillier, Göttingen, Germany
\paragraph{}
Schmerer WM (2003) Extraction of Ancient DNA. Methods Mol Biol 226: 57-61
\paragraph{}
Schmerer WM (2021a) Optimized protocol for DNA extraction from ancient skeletal human remains using Chelex-100. Int J For Med 3(1) 18-22. DOI: 10.33545/27074447.2021.v3.i1a.35
\paragraph{}
Schmerer WM (2021b) Optimized Protocol for Chelex-based Extraction of DNA from Historical Skeletal Remains and Forensic Trace Samples. Protocol Exchange, Research Square. pex-1652/v1. DOI: 10.21203/rs.3.pex-1652/v1
\paragraph{}
Schmerer WM (2022) One for All and All in One: Modified Silica Kit-based Protocol for simultaneous sample-specific Extraction of DNA from a Variety of Source Materials. J Anal Tech Res 4(2) 1: 125-129. DOI: 10.26502/jatr.28
\paragraph{}
Schmerer WM (2024) Extraction of human DNA from Soil (WMS 11) – Protocols. Published Protocols adapted to Extraction of human DNA from Soil. WM Schmerer 14Feb2024. Student instruction document, University of Wolverhampton, February 2024
\paragraph{}
Schmerer WM, Hummel S \& Herrmann B (1999) Optimized DNA extraction to improve reproducibility of short tandem repeat genotyping with highly degraded DNA as target. Electrophoresis 20(8): 1712-1716 
\paragraph{}
Schnitzer M (1978) Humic substances: Chemistry and reactions. In: Schnitzer M \& Khan SU (eds.) Soil Organic Matter. Elsevier, Amsterdam, The Netherlands, pp. 1-64
\paragraph{}
Shahzad MS, Bulbul O, Filoglu G, Cengiz M \& Cengiz S (2009) Effect of blood stained soils and time period on DNA and allele drop out using Promega 16 Powerplex® kit. Forensic Sci Int Genet Suppl 2: 161–162
\paragraph{}
Tan KH (2014) Humic Matter in Soil and the Environment: Principles and Controversies (2nd ed.) Routledge, Boca Raton, FL, USA
\paragraph{}
Thomas AE, Holben B, Dueño K \& Snow M (2019) Mitochondrial DNA Extraction from Burial Soil Samples at Incremental Distances: A Preliminary Study. J Forensic Sci 64(3): 845-851 
Tsai Y-L \& Olson BH (1991) Rapid Method for Direct Extraction of DNA from Soil and Sediments. Appl Environ Microbiol 47(4): 1070-1074
\paragraph{}
Ukalska-Jaruga A, Bejger R, Debaene G \& Smreczak B (2021) Characterization of Soil Organic Matter Individual Fractions (Fulvic Acids, Humic Acids, and Humins) by Spectroscopic and Electrochemical Techniques in Agricultural Soils. Agronomy 11(6): 1067 
\paragraph{}
Walsh PS, Metzger DA \& Higuchi R (1991) Chelex® 100 as a medium for simple extraction of DNA for PCR-based typing from forensic material. BioTechniques 10(4): 506-513
\paragraph{}
Willerslev E, Hansen AJ, Binladen J, Brand TB, Gilbert TMB, Shapiro B, Bunce M, Wiuf C, Gilichinsky DA \& Cooper A (2003) Diverse Plant and Animal Genetic Records from Holocene and Pleistocene Sediments. Science 300: 791-795
\paragraph{}
Wilson IG (1997) Inhibition and Facilitation of Nucleic Acid Amplification. Appl Environ Microbiol 63(10): 3741–3751

\end{document}